\documentclass[twocolumn,aps,prb,showpacs]{revtex4}
\usepackage{graphicx}
\usepackage{dcolumn}
\begin{document}

\title{Possible persistence of the metal-insulator transition in 
       two-dimensional systems at finite temperatures}

\author{A.~M\"obius\footnote{e-mail: a.moebius@ifw-dresden.de}}

\affiliation{Leibniz Institute for Solid State and Materials Research 
       Dresden, D-01171 Dresden, Germany}
       
\date{\today}

\begin{abstract}     
For the immediate vicinity of the metal-insulator transition (MIT), 
data on the dependence of the resistivity $\rho$ on the charge 
carrier concentration $n$ from an Si MOSFET experiment by Kravchenko et
al.\ and from an AlAs quantum well study by Papadakis and Shayegan are 
reanalyzed. In both cases, the $\rho(T = {\rm const.},n)$ curves 
for various values of the temperature $T$ seem to exhibit an offset
concerning $n$, where the related resistivity is close to 
$\mbox{h}/\mbox{e}^2$.  This offset may result from a peculiarity in 
$\rho(T = {\rm const.},n)$ indicating the MIT to be present also at 
finite $T$. More detailed experiments are imperative.
\end{abstract} 


\maketitle

The metal-insulator transition in two-dimensional (2D) systems has been 
under controversial debate for the last years. Its existence was 
denied by the localization theory by Anderson et al. \cite{Abra.etal.79} 
Thus it came as a big surprise when Kravchenko et al.\ first reported on 
a strong decrease of $\rho$ with decreasing $T$ in high mobility MOSFET 
samples. \cite{Krav.etal.94} They considered the conduction in the 
related $(T,n)$ region as metallic, an interpretation which was called 
in question in particular by Altshuler and Maslov, \cite{Alts.Masl.99} 
compare also Refs.\ \onlinecite{Krav.etal.99,Alts.Masl.99a}. A current 
review on this field is given in Ref.\ \onlinecite{Abra.etal.01}.

In the context of this debate, it is instructive to compare the usual 
approaches to data interpretation in experimental studies of the MIT in 
2D and 3D disordered systems. In the 2D case, the common intersection 
point of resistivity curves, $\rho(T = {\rm const.},n)$, measured at
various fixed temperatures, is often considered as indication of the 
MIT. \cite{Krav.etal.95} In this way, the critical concentration 
$n_{\rm c}$ is defined by 
$\mbox{d} \rho(T,n_{\rm c}) / \mbox{d} T |_{T \rightarrow 0} = 0$. 

In the 3D case, however, most publications do not pay special attention 
to the corresponding concentration value. Instead, the MIT usually 
is identified with the vanishing of the parameter $a$ in fits of the 
conductivity $\sigma(T,n = {\rm const.})$ by means of the ansatz 
$a(n) + b(n) \cdot T^p$, mostly with $p = 1/2$ or $1/3$ according to 
Refs.\ \onlinecite{Alts.Aron} and \onlinecite{News.Pepp}, respectively. 
For a typical such work see Ref.\ \onlinecite{Waff.etal}. However, this 
approach to locating the MIT suffers from inconsistencies.
\cite{Moeb.89,Moeb.etal.99}

An alternative, phenomenological approach to the data analysis for 3D
systems is based on universal features of $\sigma(T,n)$, in particular 
on a scaling law for the $T$ dependence in the hopping region,
\cite{Moeb.etal.83,Moeb.etal.85,Moeb.90a,Moeb.90b,Moeb.85} see 
also Ref.\ \onlinecite{Sara.Dai}. It yields several independent 
arguments for the MIT occurring when
$\mbox{d} \rho(T,n_{\rm c}) / \mbox{d} T |_{T \rightarrow 0} = 0$,
in analogy to the 2D case. These investigations finally lead to a 
phenomenological model for amorphous Si$_{1-x}$Cr$_x$ simultaneously 
describing both sides of the MIT, see Ref.\ \onlinecite{Moeb.90b}. 
Certain features of the model are observed also at other substances. 
\cite{Moeb.85}

This phenomenological model has an unusual feature: The MIT is expected
to persist at finite $T$, that means the $\rho(T = {\rm const.},n)$ 
curves should exhibit peculiarities as charted in Fig.\ 1. Surely, such
a sharp, but continuous MIT at finite $T$ would contradict common 
expectations. However, some indirect evidence for this hypothesis comes 
from the inspection of $\mbox{d} \sigma / \mbox{d} T$ versus $\sigma$ 
plots. \cite{Moeb.87,Rose.etal.94} Nevertheless, the situation is far 
from being clear: Detailed studies of $\rho(T = {\rm const.},n)$ in the 
$n$ region where $\mbox{d} \sigma / \mbox{d} T$ changes its sign are 
missing yet. Unfortunately, appropriate experiments as stress tuning 
\cite{Waff.etal} have focused on another concentration / conductivity
region.

\begin{figure}
\includegraphics[width=0.8\linewidth]{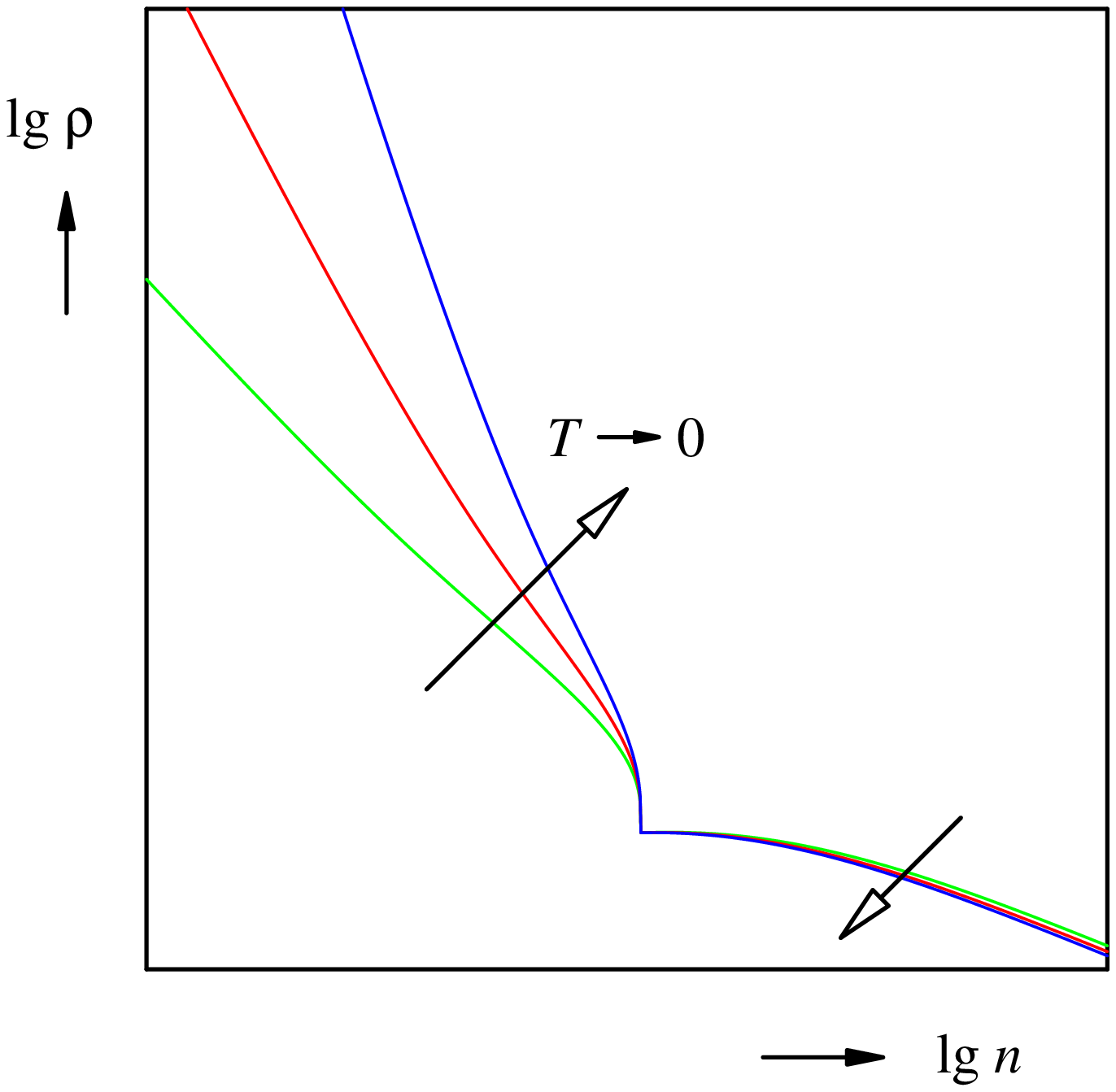}
\caption{Qualitative behavior of $\rho(T = {\rm const.},n)$ for 3D 
systems according to the phenomenological model developed in Refs.\ 
\protect{\onlinecite{Moeb.etal.83,Moeb.etal.85,Moeb.90a,Moeb.90b}},
see also p.\ 4667 in Ref.\ \protect{\onlinecite{Moeb.85}} for an 
early version.}
\end{figure}

Here I examine the question whether an analogous 
peculiarity of $\rho(T = {\rm const.},n)$ might exist in 2D systems 
--- of course, the related critical exponents would be expected to have 
other values than in the 3D case ---. For that I reconsider 
experimental data from the literature with a ``magnifying glass''.

Results on $\rho(T = {\rm const.},n)$ from the MOSFET experiment
by Kravchenko et al.\ \cite{Krav.etal.95} are replotted in Fig.\ 2 
where only the lowest four $T$ values are taken into account. The data 
were reconstructed from Fig.\ 1 of Ref.\ \onlinecite{Krav.etal.95} by 
``retranslating'' the Postscript file of the preprint version of this 
work, Ref.\ \onlinecite{Krav.etal.95a}. Note that I consider here only 
a small part of the original data: In Fig.\ 2, $n$ and $\rho$ vary by 
factors of about 1.2 and 12, respectively, whereas the original Fig.\ 1 
of Ref.\ \onlinecite{Krav.etal.95} presents variations by factors of 2 
and 4000, respectively. 

\begin{figure}
\includegraphics[width=0.8\linewidth]{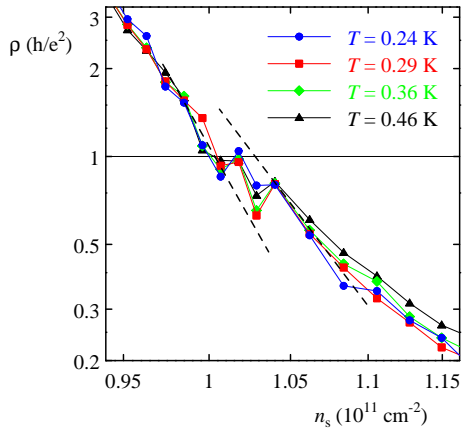}
\caption{log-log representation of $\rho(T = {\rm const.},n)$ for the 
intermediate vicinity of the MIT in a high mobility MOSFET. Data were 
obtained from Fig.\ 1 of Ref.\ \protect{\onlinecite{Krav.etal.95}}. 
The dashed lines serve only as guide to the eye.
}
\end{figure}

In analyzing these data, one notes strikingly large fluctuations
in the region around $n = 1.02 \times 10^{11} \mbox{cm}^{-2}$ for all 
$T$ values considered here. Detailed inspection uncovers a more 
important feature: $\rho(T = {\rm const.},n)$ seems to have a special 
structure there, causing an offset with respect to $n$ in the global 
variation of theses curves. This is connected with a reduction of the 
slope in the offset region. 

Remarkably, at the lower end of the related $n$ interval, $\rho$ is 
close to $\mbox{h}/\mbox{e}^2$. The accuracy of the data in Fig.\ 2 is 
not sufficient to determine a common intersection point of the 
$\rho(T = {\rm const.},n)$ curves. However, it can be stated that, if 
such a point exists, it is located within the interval $[0.95,1.05]$ 
which includes also the ``offset region''.  (Kravchenko et al.\
\cite{Krav.etal.95} regard $0.96 \times 10^{11} \mbox{cm}^{-2}$ as 
critical concentration, and claim $\rho$ to have a value of about 
$2 \, \mbox{h}/\mbox{e}^2$ there. The difference between their result 
and my cautious estimate has two reasons: Here I focus only at the 
low-$T$ region. Kravchenko et al.\ treated the possible peculiarity, 
discussed here, as random fluctuation of smooth 
$\rho(T = {\rm const.},n)$.)

Fig.\ 3 shows $\rho(T,n)$ data measured at an AlAs quantum well by 
Papadakis and Shayegan \cite{Papa.Shay}. These data were reconstructed 
by digitizing the $\rho(T,n = {\rm const.})$ dependences presented in 
Fig.\ 2 of Ref.\ \onlinecite{Papa.Shay} for $T = 0.30$, 0.45, and 
0.60 K. 

\begin{figure}
\includegraphics[width=0.8\linewidth]{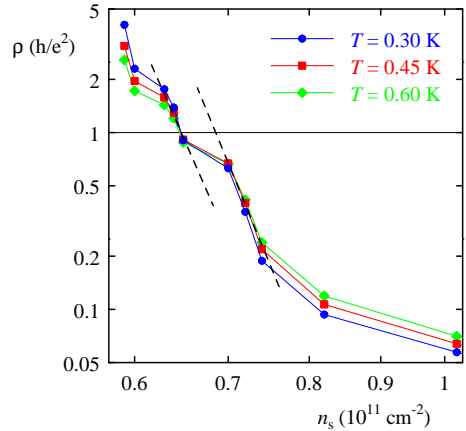}
\caption{log-log representation of $\rho(T = {\rm const.},n)$ for the 
vicinity of the MIT in an AlAs quantum well. Data were obtained from 
Fig.\ 2 of Ref.\ \protect{\onlinecite{Papa.Shay}}. The dashed lines 
are included as guide to the eye.}
\end{figure}

All $\rho(T = {\rm const.},n)$ curves given in Fig.\ 3 seem to exhibit 
a special structure between $n = 0.65  \times 10^{11} \mbox{cm}^{-2}$ 
and $0.70 \times 10^{11} \mbox{cm}^{-2}$, causing an offset in 
$\rho(T = {\rm const.},n)$ with respect to $n$. Such an offset, 
provided it is not an artifact of random errors of the $n$ values, can 
only be understood in terms of a peculiarity in 
$\rho(T = {\rm const.},n)$ there, and  / or as result of the curvature 
of $\rho(T = {\rm const.},n)$ changing the sign in this region. 

Moreover, two details of Fig.\ 3 are remarkable: At 
$n = 0.65  \times 10^{11} \mbox{cm}^{-2}$, the lower concentration 
boundary of the offset region, the values of $\rho$ are close 
to $\mbox{h}/\mbox{e}^2$. The three $\rho(T = {\rm const.},n)$ curves 
intersect each other almost at the same point, just in the region where 
this special structure seems to be located.

For both experiments, one could of course object that the emphasized
features also might arise from random deviations in the measuring values 
of $n$ and $\rho$. However, it seems to be very unlikely that all the 
following coincidences occur only by chance:

(i) These features seem to be present for all temperatures considered.

(ii) They are observed at data from two independent experiments at 
different kinds of samples, made up of different materials. 

(iii) The peculiarities have qualitatively the same form in both data 
sets.

(iv) They occur in both cases in the same resistivity region, just 
below $\mbox{h}/\mbox{e}^2$.

(v) They are observed in both cases close to the common intersection 
point of the $\rho(T = {\rm const.},n)$ curve set. 

It has to be mentioned that similar hints to a possible peculiarity 
could not be found in data from other related publications, e.g.\ 
Ref.\ \onlinecite{Puda.etal.01}. However, this is not a strong 
objection to the above interpretation since various kinds of 
inhomogeneities may wash out the comparably small effect considered 
here.

Concluding, Figs.\ 2 and 3 together, in particular the offsets in both 
curve sets, suggest that, for a certain range of finite $T$,  
$\rho(T = {\rm const.},n)$ may exhibit a peculiarity close to 
$\rho = \mbox{h}/\mbox{e}^2$. On the one hand, this would be a direct 
fingerprint of the still nowadays controversial MIT in 2D systems. On 
the other hand, it would indicate the persistence of the $T = 0$ 
phenomenon MIT at finite $T$, in contradiction to common expectations. 
Further, more detailed measurements are urgently needed to clarify this 
fundamental problem.

\begin{acknowledgments}
I'm deeply indebted to C.J.\ Adkins for motivating me to compare 
experiments on the MIT in 2D and 3D disordered systems. Critical 
discussions with M.\ Richter and G.\ Paasch have been very helpful.
\end{acknowledgments}

\end{document}